\documentclass[pre,twocolumn,superscriptaddress,showpacs,floatfix]{revtex4}
\usepackage{graphicx}
\usepackage{epsfig}
\usepackage{amssymb}
\usepackage{mathrsfs}
\usepackage{amsmath}

\begin{document}

\title{Relations between the diffusion anomaly and cooperative rearranging regions 
in a hydrophobically nanoconfined water monolayer}
\date{\today}

\author{Francisco de los Santos}
\affiliation{Departamento de Electromagnetismo y F{\'\i}sica de la
Materia, Universidad de Granada, Fuentenueva s/n, 18071 Granada,
Spain}
\author{Giancarlo Franzese}
\affiliation{Departament de Fisica Fonamental,
Universitat de Barcelona, Diagonal 645, 08028 Barcelona, Spain}

\begin{abstract}
We simulate liquid water between hydrophobic walls separated by
0.5~nm, to study how the diffusion constant $D_{\parallel}$ parallel
to the walls depends on the microscopic structure of water.  At low
temperature $T$, water diffusion can be associated with the number of
defects in the hydrogen bond network. However, the number of defects
solely does not account for the peculiar diffusion of water, with
maxima and minima along isotherms.  Here, we calculate a relation that 
quantitatively reproduces the behavior of
$D_{\parallel}$, focusing on the high-$T$ regime. We clarify how the
interplay between breaking of hydrogen bonds and cooperative
rearranging regions of 1~nm size gives rise to the diffusion extrema in
nanoconfined water.
\end{abstract}


\maketitle

In normal liquids the diffusion constant $D$ decreases when the
pressure $P$ increases at constant temperature $T$. Water, instead,
displays up to a 60\% increase of $D$ \cite{Prielmeier:1987kx} for
$P<200$~MPa \cite{Cooperativity}.  Based on simulations of bulk water
with detailed models, e.g.
Refs.~\cite{Reddy:1987bh,Starr:1999qf,Netz:2001tg}, and lattice
models, e.g. \cite{Girardi2007}, relations between the behavior of $D$
and the structure of water have been proposed.  For example, $D$ can
be related to the configurational entropy \cite{Scala:2000kl}, and the
minima in $D$ can be associated with a maximum in orientational order
\cite{Errington01}.  In classical molecular dynamics (MD) simulations
an increase of $P$ weakens the hydrogen
bonds (HBs), thus increasing $D$ \cite{Starr:1999qf}.  This
interpretation in terms of defects in the HB network can be extended
to negative $P$ \cite{Netz:2001tg}.  A similar qualitative conclusion
has been reached also by {\em ab initio} MD
\cite{Fernandez-Serra:2004eu}. Nevertheless, a quantitative relation
between the microscopic structure of water and the anomalous behavior
of $D$, including both extrema, is still missing.

Diffusion experiments in hydrophobic confinement are at issue.  For
water confined in carbon nanotubes (CNTs) with diameters from 5~nm to
2~nm, $D$ decreases for smaller diameters
\cite{Naguib:2004zr}. Nevertheless, other experiments reveal an
exceptionally fast transport for water confined in CNTs of about 2~nm
\cite{Holt:2006vn} and 7~nm diameter \cite{Majumder05}.

Results from models are also controversial.  MD simulations of five point 
transferable intermolecular potential (TIP5P)
water nanoconfined between hydrophobic smooth walls display anomalous
diffusion constant $D_\parallel$ parallel to the walls at lower $T$
than in bulk \cite{Kumar2005}, and no anomaly in the orthogonal
direction \cite{Han08}.  A large decrease of diffusion is found at
ambient conditions for extended simple point charge (SPC/E) water 
between two large hydrophobic
graphite-like plates for separations below 1.3~nm
\cite{Choudhury:2005ij}. Nevertheless, first--principles MD of the
same model in similar conditions show that the diffusion becomes
faster under confinement, possibly due to weaker HBs at the interface
\cite{cicero2008}. A similar controversy is reported for simulations
of water in CNTs, with diameters below 1~nm
\cite{cicero2008,Marti:2001kl,Mashl:2003tg}.

Here we study, by the Monte Carlo (MC) method, the diffusion of water in a
monolayer between two parallel hydrophobic walls, adopting an already
well-studied coarse-grained model \cite{FS2002}.  We find maxima and
minima of $D_{\parallel}$ and derive an expression that relates
$D_{\parallel}$ to water configurations, identifying cooperative
rearranging regions and a relation among structure, thermodynamics and
dynamics that clarifies the mechanisms for water diffusion.

We consider two flat hydrophobic walls at a distance of $h\simeq
0.5$~nm such that the formation of ice is inhibited
\cite{PhysRevLett.91.025502}. Adopting the natural square symmetry of
the system \cite{PhysRevLett.91.025502}, we divide the available
volume $V$ into $\mathscr{N}$ square cells, each with a volume
$v=V/\mathscr{N}$, and hydrate the system with $N\leq \mathscr{N}$
water molecules. To each cell we associate an occupation variable
$n_i=0,1$ ($i=1,2,\ldots, \mathscr{N}$) if it is vacant or occupied,
respectively.  The enthalpy of the system is
\begin{equation}
H\equiv \sum_{ij}U(r_{ij})
-JN_{\rm HB}
-J_\sigma\sum_in_i\sum_{(k,\ell)_i}\delta_{\sigma_{ik},\sigma_{i\ell}}+PV,
\label{eq1}
\end{equation}
where $r_{ij}$ is the distance between water molecules $i$ and $j$,
$U(r)\equiv \infty$ for $r<r_0\equiv 2.9$~\AA, the water van der Waals
diameter, $U(r)\equiv \epsilon_w [(r_0/r)^{12}-(r_0/r)^{6}]$ for
$r\geq r_0$ with $\epsilon_w\equiv 5.8$~kJ/mol, the isotropic
attraction energy, and $U(r)=0$ for $r>r_c = \sqrt{\mathscr{N}}/4$,
the cut-off distance.  $J\equiv 2.9$~kJ/mol is the characteristic
energy of the directional (covalent) component of the HB
\cite{Isaacs2000403}, $N_{\rm HB}\equiv\sum_{\langle i,j \rangle}n_i
n_j \delta_{\sigma_{ij},\sigma_{ji}}$ is the total number of HBs,
where the sum is over nearest-neighbor (n.n.)  molecules,
$\delta_{\sigma_{ij},\sigma_{ji}}\equiv1$ if
$\sigma_{ij}=\sigma_{ji}$, $\delta_{\sigma_{ij},\sigma_{ji}}\equiv 0$
otherwise, and the variables $\sigma_{ij}$ are defined as follows.  We
adopt a geometrical definition in which the HB breaks if
${\widehat{\rm OOH}}> 30^\circ$. Therefore, only 1/6 of the orientation
range $[0,360^\circ]$ in the OH--O plane is associated with a bonded
state.  Hence, we account for the entropy loss of molecule $i$ due to
the formation of a HB with molecule $j$ by associating to $i$ a
bonding index $\sigma_{ij}$, and to $j$ a bonding index $\sigma_{ji}$,
with both indices $\in [1,2, \ldots, q]$ with $q\equiv 6$.  For the
square symmetry, each molecule has four n.n. and four $\sigma_{ij}$,
with $q^4=6^4=1296$ possible bonding states. Therefore, when two
n.n. molecules $i$ and $j$ with $n_i n_j=1$ form a HB, their energy
and entropy decrease.

HB formation increases the average volume per molecule, because it induces
a local structure with a reduced number of n.n. as compared to close
packing \cite{soper-ricci2000}.  This effect is accounted for by an
enthalpy increase $Pv_{\rm HB}$ for each HB, where $v_{\rm
HB}/v_0\equiv 0.5$ is the average density increase from low density
ice Ih to high density ices VI and VIII, and $v_0\equiv hr_0^2$. The
total volume occupied by water is $V_w\equiv Nv+N_{\rm HB}v_{\rm HB}$.
The increase $v_{\rm HB}$ corresponds to a larger volume per molecule,
but not to a larger separation $r$ between molecules, having no effect
on $U(r)$.

In Eq.~(\ref{eq1}), $J_{\sigma}\equiv 0.29$~kJ/mol is the energy gain
for two bonding indices of the same molecule in the same state, and
accounts for the HB many-body (cooperative) interaction
\cite{Cooperativity,FS2002}, with
the sum over $(k,\ell)_i$, i.e., over each of the six different pairs
of the four $\sigma_{ij}$ of a molecule $i$.  This cooperative
interaction among HBs favors specific $T$-- and $P$--dependent values
of the probability distribution of O--O--O angles \cite{Soper08}. In
confinement, the distribution displays no fifth interstitial n.n. and
has a maximum shifted toward $90^\circ$ at low $T$ \cite{Ricci09},
consistent with the square symmetry adopted here.  The water-wall
interaction is represented by a hard-core exclusion.

We perform MC simulations in the $NPT$ ensemble for $\mathscr{N}=2500$
and $N/\mathscr{N}=0.75\%$, corresponding to $N=1875$ water
molecules. Since we allow for changes of the volume
in the direction parallel to the walls, the control parameter $P$ represents the 
pressure parallel to the walls.
We test that for $\mathscr{N}=400$ and $1600$ at the same
$75\%$ occupancy ratio there are no appreciable differences, as well
as for $75\%\leq N/\mathscr{N}\leq 90\%$.  The detailed MC algorithm
and the conversion to real units are described elsewhere
\cite{FS2002}.  We equilibrate each state point for 0.2~ms and average
over 15~ms \cite{note}.

\begin{figure}
\includegraphics[width=8cm,angle=0]{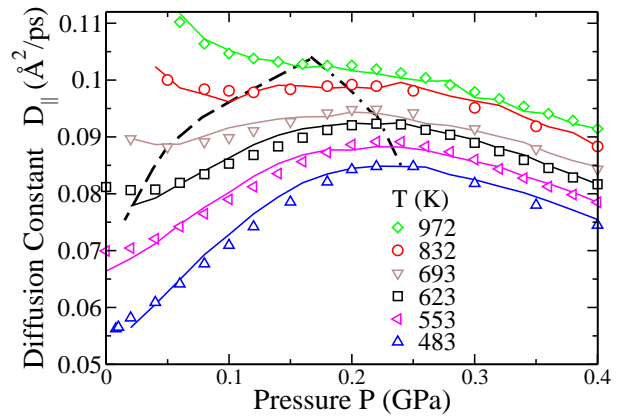}
\caption{(Color online). Diffusion coefficient $D_{\parallel}$ from MC simulations
(symbols) as a function of pressure along isotherms for (from top to
bottom) $T$ between 972 and 483~K. For $T<972$~K, $D_{\parallel}$ has
maxima $D_{\parallel}^{\rm{max}}$ at $P^{\rm{max}}$ (dotted-dashed line)
and minima $D_{\parallel}^{\rm{min}}$ at $P^{\rm{min}}$ (dashed line).
Solid lines are from Eq.~(\ref{Wnumu}) as described in the text.}
\label{diffusivity}
\end{figure}

We calculate $D_{\parallel}$ from the two-dimensional Einstein relation
\begin{equation} D_{\parallel}= \lim_{t\to \infty} \frac{\langle |{\bf
r}_i(t+t_0)-{\bf r}_i(t)|^2 \rangle}{4t}
\label{diffusion_coeff}
\end{equation}
where ${\bf r}_i(t)$ is the projection onto the plates of the position of
molecule $i$ at time $t$, and $\langle \cdot \rangle$ is the average
over all molecules $i$ and over different values of $t_0$.  We find
that $D_{\parallel}$ decreases below a maximum
$D_{\parallel}^{\rm{max}}$ for decreasing $P^{\rm{max}}\lesssim
0.2$~GPa, as in bulk water \cite{Cooperativity}, reaching a minimum
$D_{\parallel}^{\rm{min}}$ at $P^{\rm{min}}\leq P^{\rm{max}}$
(Fig.~\ref{diffusivity}).  $P^{\rm{min}}$ decreases with decreasing $T$
and eventually becomes negative.  For the high temperatures considered
here, $D_{\parallel}$ is smaller than the diffusion constant for bulk
water, qualitatively consistent with high-$T$ simulations of water in
CNTs \cite{Marti:2001kl}.

\begin{figure}
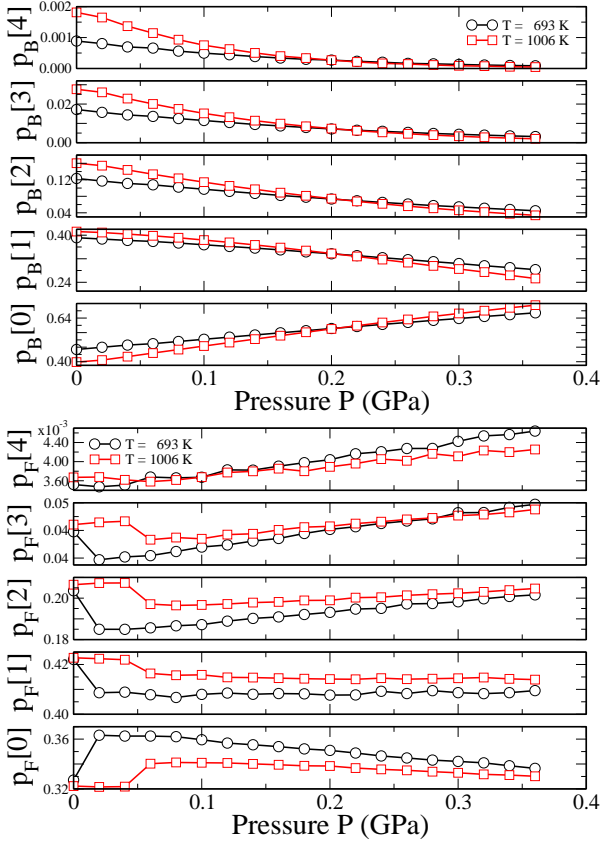

\includegraphics[width=7.9cm,angle=0]{nhb_T.8_T1.25_ur.eps}
\includegraphics[width=7.9cm,angle=0]{bp_T.8_T1.25_ur.eps}
\caption{(Color online). Microscopic configuration changes along isotherms for
temperatures 1006~K (squares) and 693~K (circles), corresponding to
the onset of anomalous $D_{\parallel}$ and to the fully developed
anomaly of $D_{\parallel}$, respectively. Top: Probability
$p_B[i]$ for a molecule to participate in $i=0, \dots, 4$ HBs.  
Bottom: Probability $p_F[i]$ for a molecule to have $i=0, \dots, 4$
free n.n. cells available for diffusion.  The discontinuity in
$p_F[i]$ corresponds to the liquid-gas first--order phase transition.
Among different panels, probability values change up to two orders of
magnitude. }
\label{neighboring}
\end{figure}

To quantitatively relate the anomalous behavior of $D_{\parallel}$
with the microscopic arrangement of water molecules, we define
$p_B[i]$ as the probability for a molecule to participate in (exactly)
$i$ HBs, with $i=0, \dots, 4$, and $p_F[i]$ as the probability for a
molecule to have (exactly) $i$ free n.n. cells available for diffusion
(Fig.~\ref{neighboring}).  By definition $\sum_i p_B[i]=\sum_i
p_F[i]=1$ and $p_d\equiv 1-p_B[4]$ is the probability to have defects
in the HB network. Because $p_B[4]$ is a monotonic function also for
temperatures at which $D_\parallel$ is non-monotonic, our result show
clearly that the network defects cannot be solely responsible for
anomalous water diffusion.

\begin{figure}
\includegraphics[width=8.75cm,angle=-0]{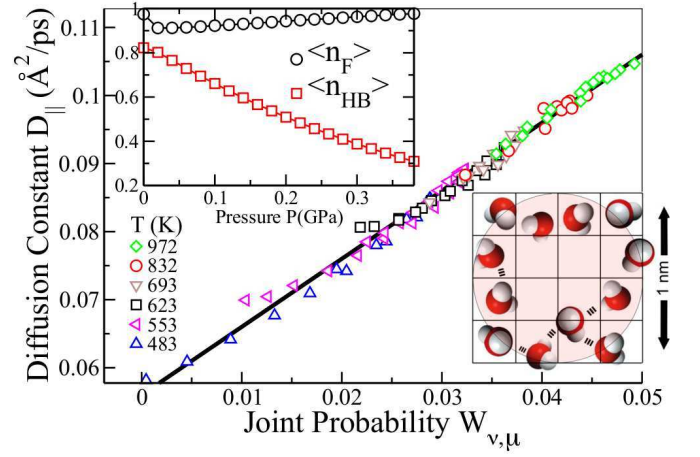}
\caption{(Color online). Upper inset: Average numbers of free n.n. cells around a molecule
$\langle n_F\rangle$ (circles) and of HBs formed by a molecule
$\langle n_{\rm HB}\rangle$ (squares) as a function of $P$ at
$T=693$~K. The discontinuity in $\langle n_F\rangle$ at low $P$
corresponds to the gas--liquid first--order phase transition. At
higher $P$, both quantities are monotonic.  Main panel: The plot of
$D_\parallel$ versus $W_{\nu,\mu}$ along the six isotherms represented
in Fig.~\ref{diffusivity} shows that $D_\parallel$ is a linear function
of $W_{\nu,\mu}$ within the numerical noise. The parameters $\nu$ and
$\mu$ change with $T$, as described in the text.  Lower inset:
Example of a cooperative (shaded) region of about 1~nm size, with $\nu=12$
molecules with n.n. free cells (not all represented in the scheme)
and $\mu~\mathcal{P}_b=5$ HBs.}
\label{averages}
\end{figure}

In particular, the probabilities $p_B[i]$ behave regularly for any $i$
(Fig.~\ref{neighboring}). The probability $p_B[0]$ of a molecule not
participating in any HB increases by increasing $P$, because of the enthalpic
cost of forming HBs at high $P$. For the same reason, the other
$p_B[i]$ to form $i>0$ HBs decrease with $P$, being vanishing small
for $i=3$ and 4.

Moreover, the behavior of $p_F[i]$ is regular within the liquid phase, with
the probability $p_F[0]$ to have 0 free n.n. cells (high density)
increasing discontinuously at the gas-liquid phase transition.
Although $p_F[0]$ decreases by increasing $P$ in the liquid phase, it
is always larger than in the gas phase. The decrease of $p_F[0]$ is
compensated by increases in $p_F[2]$, while $p_F[1]$ is almost
constant in the liquid phase and $p_F[3]$ and $p_F[4]$ are vanishing
small.  We understand the small increase of $p_F[3]$ and $p_F[4]$ as a
consequence of the phase separation between liquid and gas, favored by
the breaking of HBs due to the increase of $P$. By comparing two
different temperatures, we find that at any $P$ the higher $T$ has
lower $p_F[0]$, consistent with its lower density
(Fig.~\ref{neighboring}).

Therefore, none of the quantities $p_B[i](P)$ or $p_F[i](P)$ shows an
evident non-monotonic behavior in the liquid phase that could be
unambiguously related to the non-monotonic behavior of
$D_{\parallel}(P)$.  In the liquid phase, we also find a regular
behavior for the average number of free n.n. cells around a molecule,
$\langle n_F \rangle\equiv \sum_{f=0}^4 f~p_F[f]$ and for the average
number of HBs formed by a molecule $\langle n_{\rm HB} \rangle\equiv
\sum_{b=0}^4 b~p_B[b]$ (inset in Fig.~\ref{averages}).

We observe (Fig.~\ref{averages}) that at 693~K, for $P$ decreasing from
0.4~GPa to approximately atmospheric pressure, the average number of HBs per
molecule $\langle n_{\rm HB} \rangle$ increases from 0.3 to $\approx
0.8$, while the average number of free n.n. cells, $\langle n_F
\rangle$, around each molecule has a small decrease from 1 to
0.9. Hence, by decreasing $P$ the change in HBs is large, but not the
reduction of free volume around each molecule.

Nevertheless, diffusion in a liquid is a process that involves more
than one molecule at a time, as a consequence of the cooperative
displacement of molecules, especially at high densities \cite{DAV}.
Therefore, what is relevant is not just the available volume around a
molecule, but the volume available for diffusion $V_{DA}$ (dynamically
available volume) over a region large enough to allow for cooperative
movement of the molecules in it \cite{DAV}.

To find an analytic expression for quantifying the balance of $V_{DA}$
and broken HBs necessary for diffusion, we observe that for a
microscopic configuration with averages $\langle n_F \rangle$ and
$\langle n_{\rm HB} \rangle$ (both functions of $P$ and $T$), the
enthalpy per molecule Eq.~(\ref{eq1}) is, in mean--field approximation,
\begin{align}
H(P,T) & \equiv
H\{\langle n_F \rangle, \langle n_{\rm HB} \rangle\}  \nonumber \\
& \equiv
-\epsilon(4- \langle n_F \rangle) -(J-Pv_{\rm HB}) \langle n_{\rm HB} \rangle,
\label{Hab}
\end{align}
where we assume that the average contribution of the cooperative term
in Eq.~(\ref{eq1}) is negligible in the considered range of $P$-$T$, as
suggested by our calculations \cite{FS2002}.

Defining
$\mathcal{P}_F\equiv \langle n_F \rangle/4$
as the probability for each cell to have a n.n. cell that is
free, and
$\mathcal{P}_b\equiv \langle n_{\rm HB} \rangle/4$
as the probability for each possible HB to be formed,
the quantity
\begin{equation}
W(P,T) \equiv
\mathcal{P}_F
\mathcal{P}_b
\frac{1}{Z}
\exp[-H(P,T)/(k_BT)]
\end{equation}
is the  joint probability that, at a given $P$ and $T$,
a molecule forms a HB
and has a free n.n.  cell, with
\begin{equation}
Z\equiv \sum_{f=0}^4
p_F[f] 
\sum_{b=0}^{4-f}
p_B[b] ~
\exp[-H\{f,b\}/(k_BT)]
\end{equation}
where $H\{f,b\}$ is given by Eq.~(\ref{Hab}). The generalization of
$W(P,T)$ is 
\begin{equation}
W_{\nu,\mu}(P,T) \equiv
\mathcal{P}_F^\nu
~\mu~\mathcal{P}_b
\frac{1}{Z}
\exp[-H(P,T)/(k_BT)],
\label{Wnumu}
\end{equation}
and represents the probability, at given $P$ and $T$, of finding $\nu$
molecules with a n.n. free cell available for diffusion within a
region with $\mu~\mathcal{P}_b$ HBs (lower inset in
Fig.~\ref{averages}). If the diffusion behavior is 
dominated by the cooperative rearrangement of molecules within a given
region of the system, then $D_{\parallel}$ must be directly
proportional to $W_{\nu,\mu}$ for some values of $\nu$ and $\mu$.

A comparison with our MC simulations confirms this hypothesis
(Fig.~\ref{averages}). We find that, within the numerical noise,
$D_\parallel= \tilde{D}W_{\nu,\mu}+D_0$, where
$\tilde{D}=1 $\AA$^2$/ps.  The parameters $\nu$, $\mu$, and $D_0$ depend
only on $T$, with the first two describing the cooperative rearranging
region and the last associated with the case $\mu=0$, i. e. with the
diffusion constant at high $P$.  In particular, we find $D_0\lesssim
0.03$~\AA$^2$/ps, consistent with the observation that at very high
$P$ water recovers a normal behavior and the diffusion is extremely
small.
For the range of $P$-$T$ where we
observe the diffusion anomaly, we find $\nu=12.5\pm 0.5$ and $\mu$
decreasing from $15\pm 1$ at $T=483$~K to $\simeq 4\pm 3$ for $T\geq
623$~K.  The resulting $\nu=12.5$ suggests that for the water
monolayer between hydrophobic walls the diffusion mechanism, in the
studied $P$-$T$ range, requires a cooperative rearranging region that
extends over $\sim$3.5 water molecules, corresponding to $\sim$1 nm
(lower inset in Fig.~\ref{averages}).
The outcome for $\mu$ conveys that at lower $T$ almost all the HBs,
$\sim$18, within this cooperative rearranging region need to be broken
to allow for macroscopic diffusion, while at higher $T$ the number of
broken HBs necessary for diffusion rapidly decreases, as a consequence
of the reduced number of formed HBs (Fig.~\ref{neighboring}).

In conclusion, the result of our analysis is twofold. On the one hand,
it clarifies the mechanisms inducing the diffusion anomaly. In
particular, it shows that at constant $T$, by increasing $P$, the
number of HBs $\langle n_{\rm HB} \rangle$ decreases, implying a
decrease of the energy cost for a molecule to move and an increase of
free volume, proportional to $\langle n_F \rangle$, available for
diffusion and competing with the decrease of free volume due to the
increase of density. These competing mechanisms cause the increase of
diffusion under pressurization at low $P$ and their combined effect
reaches a maximum at a pressure above which HB formation is
unfavorable for enthalpic reasons. For the conditions considered here
(high $T$ and strong hydrophobic confinement), $D_\parallel$ is
reduced with respect to bulk.  On the other hand, our calculations
give a quantitative description of the phenomenon and show that the
competition between free volume, available for diffusion, and the
formation and breaking of HBs occurs within a cooperative region of
approximately three water molecules, or 1~nm, in the $P$-$T$ range relevant for
diffusion anomaly in a hydrophobically nanoconfined monolayer of
water.  This result recalls the finding of Ref.~\cite{donth}, where 
the size of cooperatively rearranging regions is estimated to be 
of the order 1~nm for a large number of glass forming materials close
to the glass temperature. Here, the 1-nm size is found at $T$ higher
then the bulk glass temperature, consistent with the fact that $D_\parallel$
in confinement is reduced with respect to bulk.
Finally, the fact that simulations for water confined between
graphite-like plates \cite{Choudhury:2005ij} or CNTs
\cite{Mashl:2003tg} show a large decrease of diffusion when the
confining length-scale is below 1~nm, suggests that strong
confinements of water interfere with the cooperative rearrangement
necessary for diffusion. Here the 1-nm scale emerges as a natural
length associated with the diffusion mechanism.

We acknowledge support from Junta de Andaluc\'ia (P07-FQM02725) and
MICINN (FIS2009-08451, FIS2009-10210 co-financed FEDER).

\end{document}